\documentclass[12pt,preprint]{aastex}

\shorttitle{Interstellar scintillation of PKS~1257$-$326}
\shortauthors{Bignall et al.}

\begin{document}

\title{Rapid interstellar scintillation of PKS~B1257$-$326:
  two-station pattern time delays and constraints on scattering and
  microarcsecond source structure}

\author{Hayley E. Bignall} \affil{Joint Institute for VLBI in Europe,
Postbus 2, 7990 AA Dwingeloo, the Netherlands} \email{bignall@jive.nl}

\author{Jean-Pierre Macquart\altaffilmark{1}} \affil{National Radio
  Astronomy Observatory, Socorro, NM 87801,
  USA}\email{jmacquar@nrao.edu}

\author{David L. Jauncey, James E. J. Lovell and Anastasios
K. Tzioumis} \affil{CISRO Australia Telescope National Facility, PO Box 76,
Epping, NSW 1710, Australia} \email{David.Jauncey@csiro.au,
Jim.Lovell@csiro.au, Tasso.Tzioumis@csiro.au}

\and

\author{Lucyna Kedziora-Chudczer} \affil{Institute of Astronomy,
School of Physics A28, University of Sydney, NSW 2006, Australia}
\email{lkedzior@physics.usyd.edu.au}

\altaffiltext{1}{formerly at Kapteyn Astronomical Institute,
  University of Groningen}

\begin{abstract}
  We report measurements of time delays of up to 8 minutes in the
  centimeter wavelength variability patterns of the intra-hour
  scintillating quasar PKS~1257$-$326 as observed between the VLA and
  the ATCA on three separate epochs. These time delays confirm
  interstellar scintillation as the mechanism responsible for the
  rapid variability, at the same time effectively ruling out the
  coexistence of intrinsic intra-hour variability in this source. 
  The time delays are
  combined with measurements of the annual variation in variability
  timescale exhibited by this source to determine  the characteristic
  length scale and anisotropy of the quasar's intensity scintillation
  pattern, as well as attempting to fit for the bulk velocity of the
  scattering plasma responsible for the  scintillation. 
  We find evidence for anisotropic scattering and highly elongated
  scintillation patterns at both 4.9 and 8.5 GHz, with an axial ratio
  $>10:1$, extended in a northwest direction on the sky.
  The characteristic scale of the scintillation pattern along its minor
  axis is well determined, but the high anisotropy leads to
  degenerate solutions for the scintillation velocity. 
  The decorrelation of the pattern over the baseline gives an estimate
  of the major axis length scale of the scintillation pattern. 
  We derive an upper limit on the distance to the scattering plasma of
  no more than 10\,pc.
\end{abstract}


\section{Introduction}

Interstellar scintillation (ISS) of extragalactic radio sources at
centimeter wavelengths can be used as a probe of both their
microarcsecond-scale structure and the structure of the turbulent
Galactic interstellar medium (ISM) \cite[e.g.][]{mj2002}.   However,
before scintillation can be utilized as an effective tool to probe
both the radio source structure and the ISM, it is first necessary to
establish the overall dimensions of the scintillation pattern,  and
the distance to and velocity of the medium responsible for the
scattering of the source.
 
For a few sources the variations are sufficiently rapid that it is
possible to measure a time delay in the arrival times of the pattern
of flux density variations between widely separated telescopes as the
Earth drifts through the scintillation pattern. In principle such a
time delay provides a direct determination of the instantaneous
scintillation velocity and helps ascertain the overall scale of the
scintillation pattern. When the scintillation pattern is highly
anisotropic, however, the anisotropy introduces a bias into the
observed delay, as noted by \citet{ps55} in an analysis of ionospheric
scintillations.  For many sources affected by ISS the Earth's orbital
motion causes a substantial change in the relative velocity between
the Earth and the scattering plasma which in turn changes the observed
pattern time delay.  In this case, measurements of the time delay at
various epochs during the year can be used to constrain the anisotropy
and position angle of the scintillation pattern.

PKS~1257$-$326 is a flat spectrum, radio-loud quasar at $z=1.256$
which exhibits intra-hour flux density variability at frequencies of
several GHz due to ISS \citep[][ hereafter Paper I]{big2003}.  The
source's rapid variability makes it possible to measure a precise time
delay between two widely separated telescopes, and hence to determine
its structure through measurement of its scintillation characteristics.

The recent MASIV Survey of 700 flat-spectrum radio sources
\citep{lov2003} showed that rapid variability of the type exhibited by
PKS~1257$-$326 is extremely rare.  Only two other quasars are known
that vary  on timescales short enough to allow time delay measurements
to be made.  These are PKS~0405$-$385 \citep{ked97} and J1819$+$3845
\citep{dtdb2000}.  Simultaneous observations of  PKS~0405$-$385 showed
a time delay of around 2 minutes between the Australia Telescope
Compact Array (ATCA) and the Very Large Array (VLA)
\citep{iau205:jau2000}.  Observations of J1819$+$3845 at the
Westerbork Synthesis Radio Telescope and the VLA  also showed a clear
delay between the pattern arrival times at each telescope, which
changed over the course of the observation due to the rotation of the
Earth and hence of the projected baseline vector \citep{dtdb2002,dtdb2003}.
Similar time delay techniques have been used to study ISS
of pulsars \citep[e.g.][]{lr70}, and interplanetary scintillation of
AGN \citep[e.g.][]{ck78}.

In the regime of weak scattering, relevant to the scintillation of
PKS~1257$-$326 above $\sim 5\,$GHz, the length scale of the
scintillation pattern is related to the Fresnel scale, $r_{\rm F} =
\sqrt{cL/(2\pi\nu)}$, where $\nu$ is the observing frequency and $L$
is the distance to the scattering medium which is assumed to be
confined to a plane.  For the case of isotropic scattering, a source
of angular diameter $\theta_{\rm S}$ larger than the angular Fresnel
scale, $\theta_{\rm F}=r_{\rm F}/L$, increases the scale length of the
scintillation pattern by a factor $\sim \theta_{\rm S}/\theta_{\rm F}$
\citep[e.g.][]{nar92}.  Thus if the angular size of PKS~1257$-$326 is
several times larger than $\theta_{\rm F}$, then the scale of the
scintillation pattern should be determined chiefly by source
structure. Assuming the same scattering material is responsible for
the scintillation at different frequencies, then the time delay is
independent of observing frequency {\em unless} source structure
changes with frequency to produce anisotropy in the scintillation
pattern which is significantly different at different frequencies. If
there is significant anisotropy in the scattering medium, however,
then the medium may have a more dominant influence on the structure of
the scintillation pattern. 

We present here the results of simultaneous ATCA and VLA observations
of PKS~1257$-$326, carried out during 2002 and early 2003.  These are
combined with measurements of the characteristic timescale of
variability from ATCA data collected over a three-year period, in
order to determine the scintillation velocity and dimensions of the
pattern.


\section{Observations and Data Reduction}

\subsection{Observing Strategy}

PKS~1257$-$326 was observed simultaneously at 4.9 and 8.5 GHz with the
VLA and the ATCA in three separate epochs, 2002 May 13 and 14, 2003
January 9 and 10, and 2003 March 6 and 7.  The accuracy with which the
time delay can be measured depends on the characteristic timescale
and on the signal to noise ratio of the observations;  to detect a time
delay, significant changes must be observable within periods of order
1 minute.  The observing sessions were therefore chosen to be during
the part of the year when the variations are most rapid, between
December and May (see Paper I).  The observations were performed on
two consecutive days in each epoch in order to measure time delays for
two independent samples of the scintillation pattern in the same part
of the annual cycle. Unfortunately in 2003 January a problem occurred
with the VLA and data from only one day, January 10, were usable.

Observations were conducted simultaneously with bands centered at 4.86
and 8.46~GHz (nominally 4.9 and 8.5 GHz in the text hereafter).  The
wideband feedhorns at the ATCA allow simultaneous observations at the
two frequencies, using the standard 128~MHz continuum bandwidth,
correlated in 32 overlapping 8~MHz channels.  The VLA was divided into
two sub-arrays of 13 and 14 antennas in order to observe both
frequencies simultaneously. Each VLA subarray observed two contiguous
50~MHz bands on either side of the center frequency. Spectral channels
were later discarded from the band edges in the ATCA data to match the
frequency coverage exactly with the 100~MHz bandwidth of the VLA.

During the period of common visibility which lasts 2.8 hours,
PKS~1257$-$326 is at low elevation at both telescopes, where the
effects of atmospheric opacity, pointing and antenna gain changes are
most severe. Since the reliability of any time delay measurement
depends critically on the flux density measurement accuracy, the
unresolved $\sim 2$~Jy compact source, PKS~1255$-$316, located
$1.2^{\circ}$ from PKS~1257$-$326, was observed at both telescopes for
calibration of these effects.  One-minute observations of
PKS~1255$-$316 were interspersed between 15-minute observations of
PKS~1257$-$326.  PKS~1255$-$316 is a good calibrator as it is
unresolved at both telescopes (more than 99\% of the total flux
density is unresolved with  the VLA in A-array) and shows no intraday
changes larger than $\sim 1$\%, although it is variable on timescales
of several months. The weather was fine at both telescope sites for
all of the observations.

\subsection{Data reduction}

\begin{figure}
\epsscale{.64} \plotone{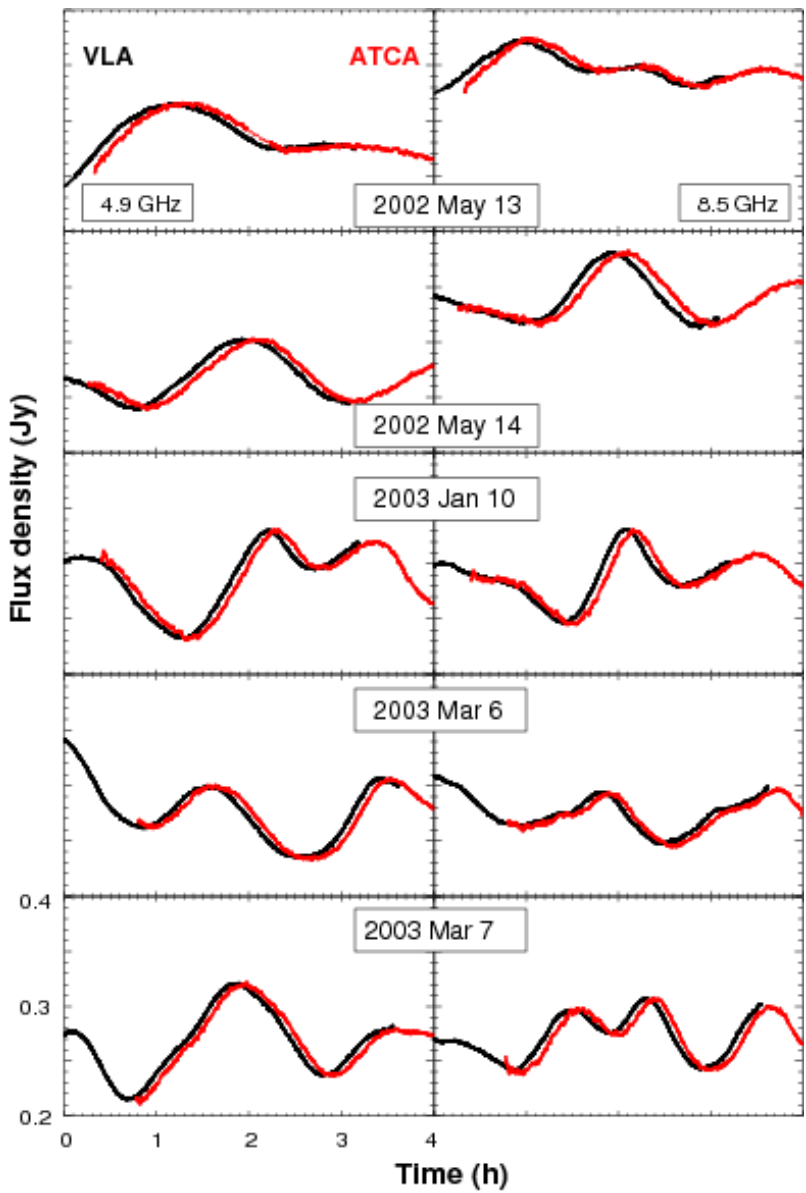}
\caption{Simultaneous observations of PKS~1257$-$326 at the VLA and
  ATCA on 5 different days. 4.9\,GHz data are shown on the left and
8.5\,GHz data on the right. All plots are on the same scale, showing a
4-hour time range, and flux density range from 0.2 to 0.4~Jy. The time
ranges shown are 05--09h UT for the data from May, 13--17h UT for the
January data, and 09--13h UT for the March data.  The flux densities
are measured from the visibilities averaged over all baselines of each
array, and are plotted here after boxcar smoothing with a width of
50\,s. The data were calibrated using observations of the nearby
compact source PKS~1255$-$316, and a model of arcsecond-scale extended
structure has been subtracted from each of the PKS~1257$-$326
datasets.}
\label{fig-alltimedel}
\end{figure}

Here we describe the procedure used for data reduction; in
\S\ref{sec-err} we estimate the resultant uncertainties in flux
density measurement accuracy. Initial data processing was performed
using standard tasks in the AIPS\footnote{The Astronomical Image
Processing System (AIPS) is developed and distributed by the National
Radio Astronomy Observatory.} and MIRIAD \citep{stw95} software
packages. AIPS was used for calibration of the VLA data, and MIRIAD
for calibration of the ATCA data.  A clear time delay, with the VLA
light curves leading those of the ATCA, is evident between the
variability patterns seen in Figure~\ref{fig-alltimedel}  at each
telescope, and could be seen in the data even before off-line
calibration.  To measure the delay accurately, antenna gains including
pointing- and time-dependent errors were determined for both
telescopes at 15 minute intervals using the calibration source
PKS~1255$-$316.

For the VLA data, initial gain curves are applied before solving for
additional gain or opacity changes using PKS~1255$-$316. For the ATCA,
we did not apply gain-elevation corrections separately but solved for
all effective gain changes simultaneously assuming a constant flux
density for PKS~1255$-$316. For both telescopes, antenna gains vary
over the course of the observations by a total of typically 3\% at
8.5\,GHz and $\sim 1.5$\% at 4.9\,GHz -- here we lump together effects
of opacity, pointing and gain-elevation changes as contributing to the
overall effective ``gain'' variations.  These variations appear to be
tracked very well by the calibration source observations at 15-minute
intervals, as outlined below in \S\ref{sec-err}. The flux density of
PKS~1255$-$316 was determined from the ATCA data and set to be
identical at both telescopes. In 2002 May the measured flux densities
of PKS~1255$-$316 were 2.4\,Jy at 4.9\,GHz and 2.3\,Jy at 8.5\,GHz,
which decreased to 2.2\,Jy at 4.9\,GHz and 1.9\,Jy at 8.5\,GHz in 2003
March. The ATCA flux density scale is tied to the primary calibrator
PKS~1934$-$638. However it is the accuracy of the relative calibration
between the two telescopes which is the principal concern for this
experiment, rather than the overall flux density scale.

\begin{figure}
\epsscale{1.0}\plottwo{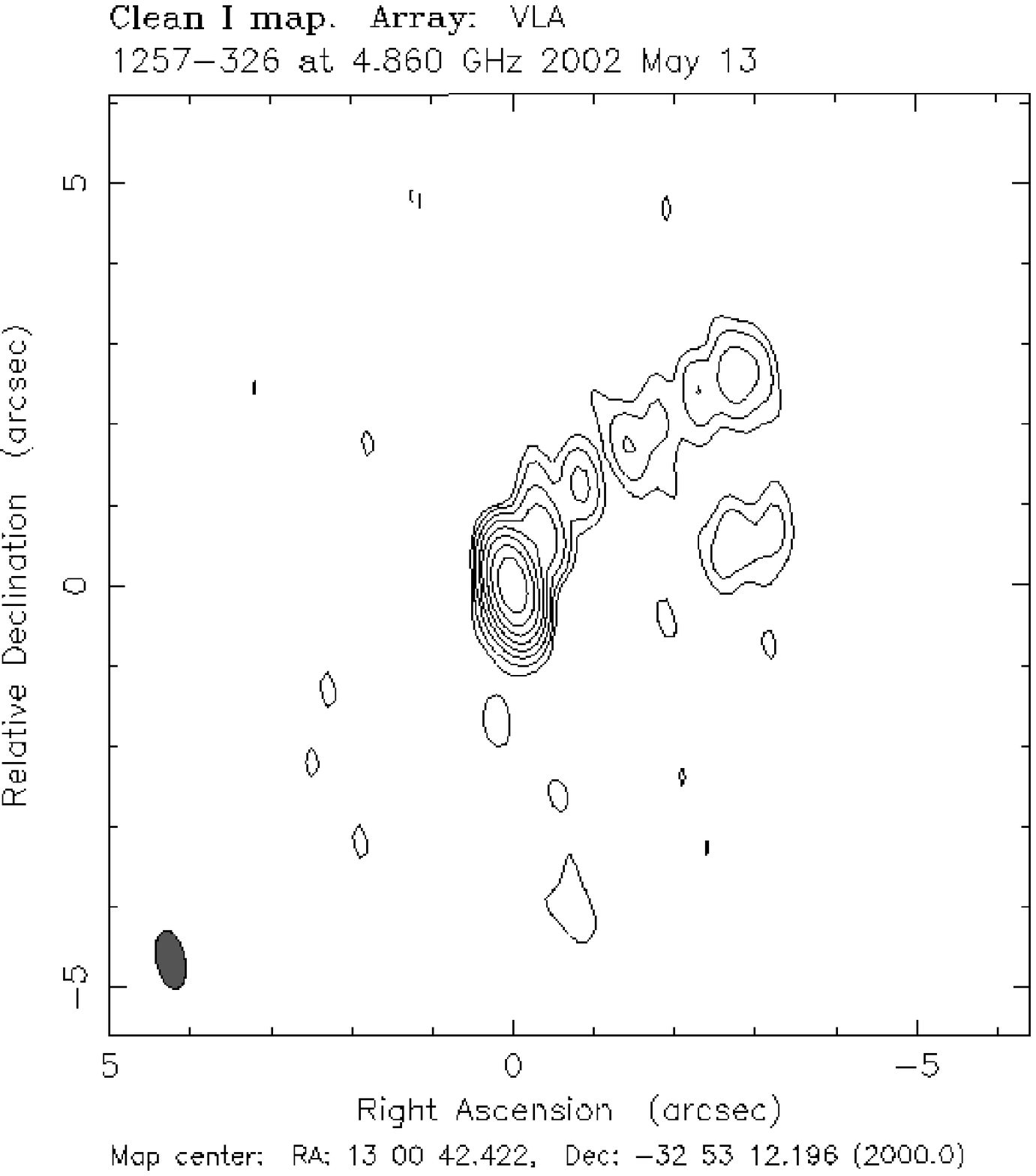}{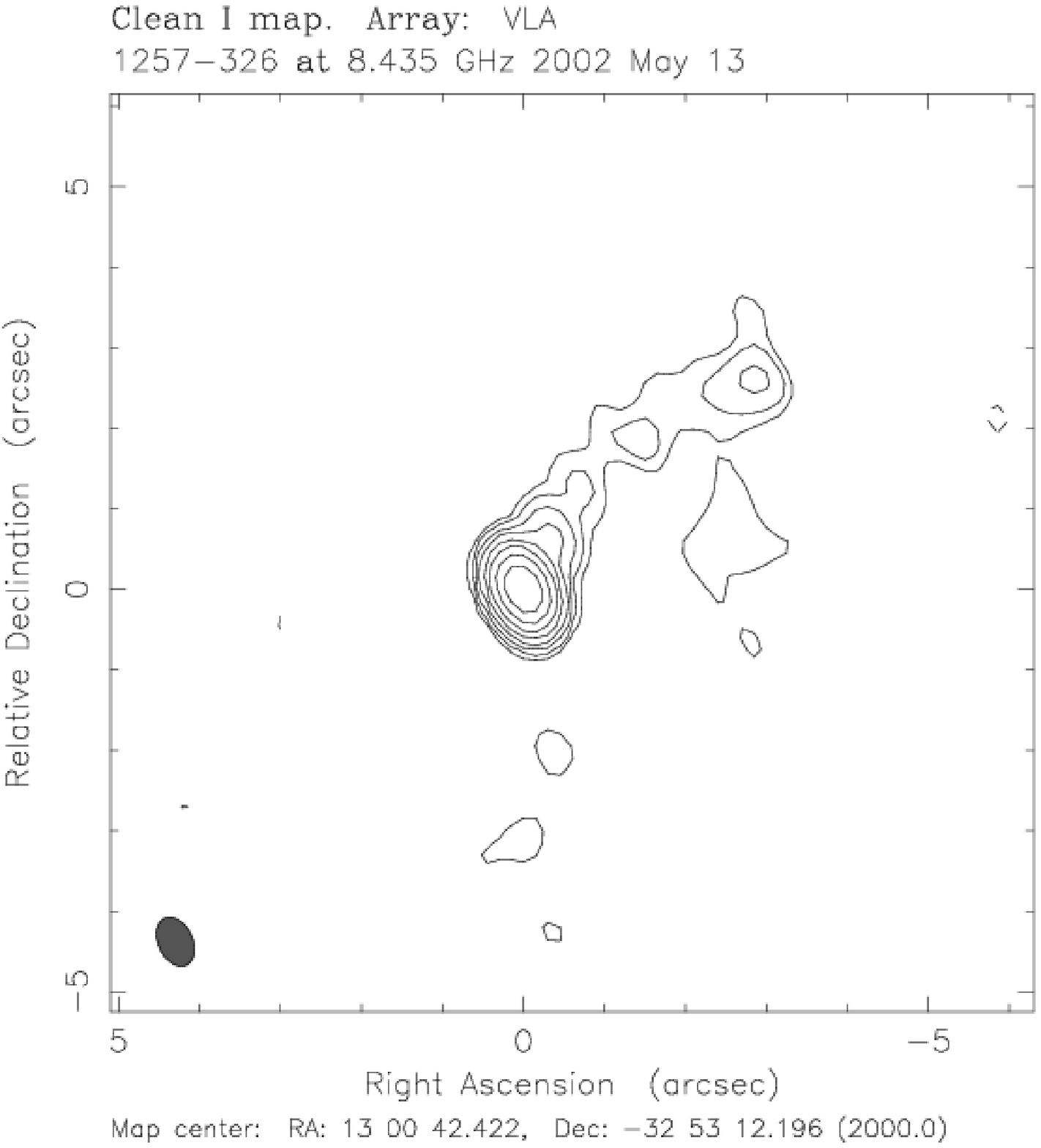}
\caption{Images showing the arcsecond-scale extended structure in
  PKS~1257$-$326, made from VLA data after subtraction of a
  time-variable point source model from the $(u,v)$ data at both 4.9
  and 8.5~GHz. In each case the subtracted flux density was the
  variable part plus the core such that the resulting map peak is
  50\,mJy/beam, with contours shown here at $-1,1,2,4,...128 \times
  0.2$\,mJy/beam at both frequencies. In the 8.5\,GHz map displayed
  here, long baselines have been weighted down with $(u,v)$ tapering
  in order to show clearly the similarity of the unusual extended
  structure at both frequencies. At each frequency a model of the
  extended structure (i.e.\ {\sc clean} components outside the
  brightest pixel) was subtracted from all the original calibrated
  $(u,v)$ data, in order to accurately compare the flux density of the
  scintillating ``core'' component measured at each array.}
\label{fig-im}
\end{figure}

At 4.9\,GHz 5\% of the total flux density of PKS~1257$-$326 is
contained in steep-spectrum components extended over several
arcseconds to the north-west of the core,  with a broad component
offset to the south of the main jet. The same structure is also seen
at 8.5\,GHz, as shown in the VLA images in Figure~\ref{fig-im}.  In
order to accurately compare the flux densities of the scintillating
component measured at both telescopes, it is necessary to subtract the
contribution of the extended emission, which varies with changing
$(u,v)$ coverage for each baseline of each array. We made images of
PKS~1257$-$326 from the VLA data obtained during 2002 May, using the
Difmap software package \citep{shep97}. The VLA at this time was in
reconfiguration between the ``A'' and ``BnA'' configurations, which
provided  the highest resolution images.  The scintillation in PKS
1257$-$326 is sufficiently rapid to introduce artefacts into an image
made from a typical  synthesis observation. The VLA has good,
two-dimensional instantaneous $(u,v)$ coverage compared with the
linear East-West ATCA configuration, allowing an excellent model of
the source to be obtained from a single VLA scan of a few
minutes. Deeper images of the extended structure were made using 4
hours of VLA data from each subarray, 
after first subtracting a point source model of
variable flux density from each 30 seconds of the data in the
visibility domain; the resultant images are displayed in
Figure~\ref{fig-im}.

The resultant model {\sc clean} components, except for the central
bright point source component, were subtracted from the PKS~1257$-$326
visibilities for both telescopes, using AIPS task {\sc uvsub}. This
corresponded to a total of 16~mJy being subtracted from the 4.9\,GHz
data, and 7~mJy from the 8.5\,GHz data. The resultant visibilities
thus represent {\em only} the component which is unresolved at both
telescopes, which contains the scintillating component.   For all
datasets, the data after subtraction of the model are consistent with
a point source, and we can conclude that our VLA images are not
missing any significant larger scale source structure. The resultant
visibilities were written to FITS files and then read into MIRIAD, at
which time a decrement of 32 seconds was applied to the VLA data time
stamps to convert  from IAT to UTC.  Flux densities, averaged over the
frequency band and over all baselines for each telescope,  were then
written out as a function of UT for further analysis, along with the
rms scatter about the mean for each data point.  At this stage no
further time averaging was applied beyond the integration times used
by each telescope's real-time correlator, 3.3\,s at the VLA and 10\,s
at the ATCA. The effect of smoothing and interpolation using a range
of time intervals was tested later in the time delay fitting process
described in \S\ref{sec-results}.

\subsection{Flux density measurement uncertainties}
\label{sec-err}

If each array sees an identical scintillation pattern, then the
expected rms difference $\sigma_{\rm diff}$ between flux density
measurements at each array, after correcting for the time delay, is a
combination of effects of antenna amplitude gain errors, a
contribution $x$\,mJy due to errors in the model of extended
structure,  and the thermal noise for each array $\epsilon$
mJy. Antenna gain errors contribute an overall uncertainty which is a
fraction of the total flux density, $gS$.  The various errors for each
array are independent and hence add in quadrature.  Thus
\begin{eqnarray} 
\sigma^2_{\rm diff} = \sigma^2_{\rm V} + \sigma^2_{\rm A} = 
\epsilon_{\rm V}^2 + \epsilon_{\rm A}^2 + (g_{\rm V}S)^2 + 
(g_{\rm A}S)^2 + 2x^2,
\label{errdiff}
\end{eqnarray}
where the subscripts V and A denote errors for the VLA and ATCA
respectively.

Based on the amplitude gain corrections found for the calibration
source PKS~1255$-$316, we estimate that our calibration procedure is
sufficient to calibrate the {\em target} source visibility amplitudes
averaged over all baselines to an accuracy of $g \sim 0.1$\% at
4.9\,GHz and $g \sim 0.25$\% at 8.5\,GHz at both telescopes, where the
flux density scale is tied to PKS~1255$-$316.  Errors in the
subtracted model, and their effect on the averaged visibilities for
the various array configurations, are difficult to quantify, but based
on map residuals we estimate that these contribute $x \sim 0.5$\,mJy
to the overall uncertainties. For an averaging interval of 50\,s,
corresponding to our chosen smoothing interval for the final fits, the
rms scatter due to thermal noise is typically 0.45\,mJy for the VLA,
and 1.0\,mJy for the ATCA, at both frequencies and for all observed
epochs. For both arrays, having Cassegrain-focus antennas, the system
temperature, and hence the rms measurement noise, increases toward low
elevations, as do the fractional errors due to opacity and
gain-elevation effects. As each array points toward lower elevations
at opposite ends of the observation, the overall uncertainties are
relatively constant, with the measurement accuracy being slightly
higher in the middle of the observation.

At 4.9\,GHz the expected contribution of measurement errors (in mJy)
to $\sigma^2_{\rm diff}$ for an averaging interval of 50\,s and
average flux density $\bar{S} \approx 300$\,mJy is thus
$\sigma^2_{\rm diff}{\rm(4.9~GHz)} \approx 0.45^2 + 1.0^2 +
2(0.001\bar{S})^2 + 2(0.5^2)$ giving $\sigma_{\rm diff} \approx
1.4$\,mJy.
The largest contribution to the errors at 4.9\,GHz is the thermal
noise in the ATCA data.
At 8.5\,GHz all error contributions are approximately the same as at
4.9\,GHz except for the fractional error term which is estimated to be
$2(0.0025\bar{S})^2$. Correspondingly, $\sigma_{\rm diff} \approx
1.7$\,mJy at the higher frequency for $\bar{S}=300$\,mJy and a
smoothing interval of 50\,s.


\section{Results}
\label{sec-results}
Figure~\ref{fig-alltimedel} shows the flux densities of PKS~1257$-$326
measured simultaneously at the ATCA and VLA on the 5 different days at
the two frequencies, after subtraction of the model of extended source
structure from each dataset. Data are displayed with 50\,s boxcar
smoothing.   The  scintillation patterns seen at each telescope appear
virtually identical, but clearly displaced in time with the VLA light
curve leading that of the ATCA.  Data from all epochs and both
frequencies show the same effect, with the pattern arriving first at
the VLA and then some minutes later at the ATCA.

There are two questions that we wish to address with the present
data:

(i) What are the observed time delays and what are their
uncertainties?

(ii) How similar are the scintillation patterns seen at the two
telescopes?

The measured time delays together with the observed annual cycle data
(Paper I) can be used to constrain the  properties of the scattering
medium and its peculiar velocity.  From (ii) we can also determine
lower limits on the characteristic length scale of the major axis of
the scintillation pattern. In this section we present the results of
fitting to the time delay data. \S\ref{sec-analysis} presents a
combined analysis of the time delay and annual cycle data to determine
the scintillation parameters.
 
The variability patterns can be compared directly to search for any
differences by fitting first for a time offset and then subtracting
the two patterns. We discuss below the possibility of a small change
in the pattern arrival time delay during the course of each
observation.  However, to begin with, a constant time offset was
estimated by minimizing the quantity  $\chi^2 = \sum_{i}[S_{V,i} -
S_{A,i}(\Delta t)]^2/\sigma_{\rm diff}^2$,  where trial offsets
$\Delta t$ are  applied to the ATCA data, and $S_{V,i}$ and $S_{A,i}$
are the VLA and ATCA flux density measurements, respectively,
interpolated onto the same grid in time. For each of the 10 datasets,
$\Delta t$ was allowed to vary in steps of 1\,s between 0 and
1000\,s. For the final results presented here we applied boxcar
smoothing to the data with a width of 50\,s, which corresponds to the
interval where the average change in flux density, $\sim 1$\,mJy, is
approximately equal to the rms measurement noise.  A range of
different smoothing and sampling intervals for the light curves were
tried in order to test the robustness of the fitted time delay. Using
a smaller window for smoothing (30\,s or 10\,s) made no significant
difference to the best fit time delay, but resulted in somewhat
noisier residuals around the best fit delay. We also tried tests such
as dropping low elevation data, e.g. the first and last 20 minutes of
each dataset, but this also made essentially no difference to the
fitted time delays. This gives us considerable confidence in the
robust quality of the data and in our estimation procedures.

\begin{figure}
\epsscale{0.9} \plotone{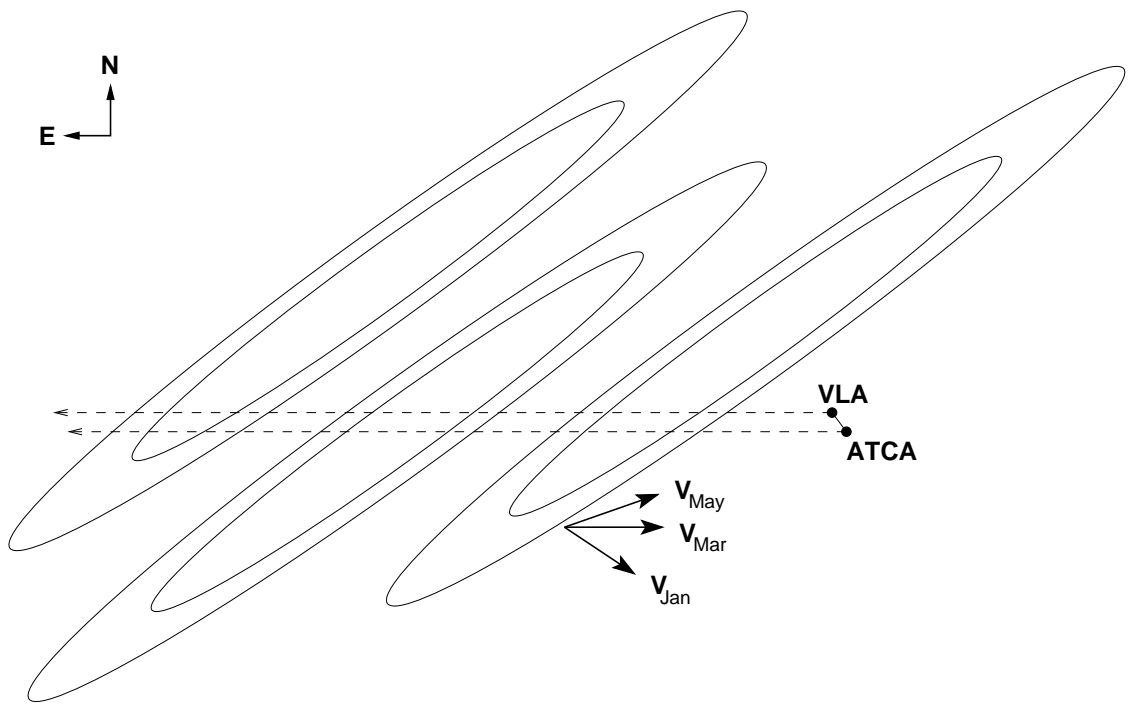}
\caption{Sketch illustrating the geometry of the time delay model (not to
  scale, but the illustrated anisotropy, angles and velocities shown
  are approximately consistent with our preferred solution).  The
  ellipses represent contours of constant intensity in the
  scintillation pattern, assumed to be ``frozen-in'' over a timescale
  of at least several minutes. Arrows indicate the direction of the
  velocity {\bf v} at which the scintillation pattern passes over the
  telescopes for each observed epoch. The straight dashed lines show
  the cut through the scintillation pattern observed at each telescope
  for the direction of ${\bf v}_{\rm Mar}$ shown.}
\label{fig-geom}
\end{figure}

Is fitting for a constant time delay the best we can do?  The
projected VLA-ATCA baseline rotates by almost $20^{\circ}$ from the
beginning to the end of the observation. As the baseline is close to
one Earth diameter there is little change in projected baseline length
over the observations. The time delay will be influenced by the
structure of the scintillation pattern and its direction of motion
relative to the baseline. One can picture this using the geometry
illustrated in Figure~\ref{fig-geom}, where an elliptical
scintillation pattern is assumed to represent the average or
characteristic scintillation scale. In reality the scintillation
pattern is stochastic and it is important to bear in mind that the
small sample of the scintillation pattern obtained in a single time
delay measurement may not be representative of the average, which was
the principle reason for repeating the time delay measurement on
consecutive days. We have tested for a change in delay within each
observation by splitting the dataset and fitting a delay to each
segment, but do not detect any significant differences in delay
between subsets. We therefore conclude that fitting a constant time
delay for each dataset is justified. However, the assumption of a
constant delay is not exactly correct, which will increase the scatter
in the residuals to some degree. Moreover, the assumption that the ISM
is well characterised by a single velocity may not be strictly
correct. Velocity structure in the medium would cause some
decorrelation in the pattern as it drifts, and hence would increase
the residual differences.

For the case of J1819+3845 \citep{dtdb2002}, a large change and
reversal in sign of the time delay was seen over the course of the two
simultaneous VLA/WSRT observations of this source, due to the large
angle through which the projected baseline rotated. For PKS~1257$-$326
we observed at different times of the year in order to measure the
time delay for different projections of the scintillation velocity
along the baseline.

For the final time delay  fitting we used a 50\,s averaging interval,
giving approximately 180 independent time intervals in the $\sim 2.5$
hours of data overlap.  We have also performed the fitting after
normalizing the flux densities by their mean values in the overlapping
time range. Normalizing effectively removes constant fractional
errors. Small calibration offsets on the level of a few tenths of a
percent were found in some datasets -- in particular the March 8.5\,GHz
data -- after processing.  Using either the measured flux densities or
the flux densities normalized by the mean makes no significant
difference to the best fit time delays or their uncertainties. 
However, using the  flux density errors estimated in \S\ref{sec-err}, the
minimum $\chi^2$ values in all cases indicate a poor fit of the model
to the data. The larger $\chi^2$ values at 8.5\,GHz suggest that atmospheric
effects may add to the uncertainties such that $\sigma_{\rm diff}$ is
slightly underestimated for most datasets. Alternatively,  
additional pattern differences may also result from either or both of (i)
spatial decorrelation resulting from each telescope seeing a different
part of the pattern, since the velocity has a component perpendicular
to the baseline, or (ii) some temporal rearrangement of the pattern on
the timescale of the delay, or equivalently, decorrelation resulting
from velocity structure in the medium, so that the observations are
not well modeled by a single time delay.  

Despite the large values, $\chi^2$ in all cases has a well-defined
minimum, and therefore we have assumed it is valid to estimate the errors
from the change in $\chi^2$ about its minimum. We calculate 1$\sigma$
uncertainties using the standard definition. The errors determined
this way  are approximately consistent with the internal scatter for
the epochs which have four datasets (March and May). The similarity of
the time delays on consecutive days gives us confidence that the
measured delay from each observation is representative of the overall
scintillation parameters, despite the small number of maxima and
minima observed in each dataset. Also, we have combined the data to
estimate a single value for the time delay in each epoch. 
Table~\ref{tab-dt} shows all best fit values of $\Delta t$, along with
estimated 1$\sigma$ errors.

\begin{deluxetable}{ccccccccc}
\tablecolumns{9}
\tablecaption{Best fit time delays and correlation coefficients\label{tab-dt}}
\tablehead{
\colhead{Obs date} & \colhead{$\Delta t_{4.9}$} &
\colhead{$\sigma_{4.9}$} & \colhead{$\chi^2_{4.9}$} &
\colhead{$\rho_{4.9}$} & \colhead{$\Delta t_{8.5}$} &
\colhead{$\sigma_{8.5}$} & \colhead{$\chi^2_{8.5}$} &
\colhead{$\rho_{8.5}$}\\
\colhead{} & \colhead{(s)} & \colhead{(s)} & \colhead{} & \colhead{}
  & \colhead{(s)} & \colhead{(s)} & \colhead{} & \colhead{}}
\startdata
2002 May 13 & 495 & 20 & 1.16 & 0.99936(30) & 496 & 27 & 1.40 & 0.99638(50)\\
2002 May 14 & 468 & 19 & 1.55 & 0.99863(20) & 475 & 24 & 2.05 & 0.99591(40)\\
\cutinhead{May combined:~~$\Delta t$ = 483 $\pm$ 15 s,~~$\chi^2$ = 1.56}
2003 Jan 10 & 340 & 12 & 1.20 & 0.99979(05) & 330 & 15 & 1.55 & 0.99872(15)\\
\cutinhead{Jan combined:~~$\Delta t$ = 333 $\pm$ 12 s,~~ $\chi^2$ = 1.38}
2003 Mar 06 & 333 & 14 & 1.10 & 0.99944(10) & 305 & 24 & 1.10 & 0.99951(50) \\
2003 Mar 07 & 313 & 10 & 1.12 & 0.99977(05) & 314 & 14 & 1.58 & 0.99897(04)\\
\hline\\[-3.5mm]
\multicolumn{9}{c}{Mar combined:~~$\Delta t$ = 318 $\pm$ 10 s,~~$\chi^2$ = 1.24}
\enddata
\tablecomments{Time delays $\Delta t$ and uncertainties $\sigma$ at
  each frequency are given, along with minimum reduced
  $\chi^2$ values and correlation coefficients, $\rho$, for the delay-corrected
  data as discussed in \S\ref{sec-results}. Uncertainties in the last
  two digits of $\rho$ are indicated in parentheses. The $\chi^2$ for each
  dataset has typically 180 degrees of freedom. Fits to the combined
  data from each epoch are also given.}
\end{deluxetable}

The rms scatter in the delayed-difference distributions is generally
not much larger than expected due to measurement uncertainties,
however there is some noticeable variation between different epochs,
with the difference light curves from May showing larger scatter than
the other epochs. 

In order to address the question of whether there is any decorrelation
of the scintillation pattern between the two telescopes, we have
calculated the linear Pearson correlation coefficient, $\rho$, for
each of the datasets after correcting for the best-fit delay. We have
also corrected $\rho$ for the instrumental noise. In all cases $\rho >
0.995$ after applying the time delay correction, while the values
before correcting for the delay range between 0.91 and 0.97. 
Values of $\rho$ for the delay-corrected data are shown in
Table~\ref{tab-dt}. These were calculated after applying 90\,s boxcar
smoothing to the data. For both days of observation in May, the correlation
coefficients at each frequency are somewhat lower than for the other
epochs. This effect is more significant at the higher frequency. As
discussed in \S\ref{sec-analysis}, a larger decorrelation in May would  
be expected based on the geometry of annual cycle for PKS 1257$-$326.

We have considered the possibility that a larger decorrelation in May
might be due to the change in time delay over the course of the
observation producing a larger variance in the residual data from this
epoch. Our models imply the time delay may change by up to $\sim 10$\%
over the course of the observation due to baseline rotation.  As a
check we repeated the calculation after applying the predicted time
delays for every 10 minutes of data using models from
\S\ref{sec-analysis} rather than a constant time delay. However this
made no significant difference to the statistics of the residuals or
the correlation coefficients.

From Figure~\ref{fig-geom} it can be inferred that when the
scintillation velocity is exactly parallel to the baseline, both
telescopes will see an identical pattern provided there is no temporal
decorrelation of the pattern on the timescale of the
delay. Alternatively, any difference between the patterns seen in
this case would be evidence of temporal decorrelation, which would most likely
be a result of velocity shear in the ISM.  The fact that the time
delays are well determined to an accuracy of 3--5\% implies that the amount of
velocity shear in the scattering medium is not significant on this
level. When the pattern drifts at some angle to the baseline, then the second
telescope sees a part of the pattern displaced with respect to the
first, as illustrated by the straight dashed lines in
Figure~\ref{fig-geom}. The amount of spatial decorrelation is also
influenced by the degree of anisotropy in the pattern and the angle
between the scintillation velocity and the direction of
anisotropy. This has been taken into account in the analysis presented
in \S\ref{sec-analysis}.

Finally, it is worth noting that through these accurate flux density and time
delay measurements, we can state that PKS~1257$-$326 shows no intrinsic
variations at either frequency greater than $\sim 2$\,mJy rms on a timescale
of 2.5 hours. The time delay observations are of course not sensitive to
variations on a timescale longer than the overlap period.

\subsection{Annual cycle in characteristic timescale}
\label{sec-ac}
In Paper I we presented measurements of the characteristic timescale
of variability for PKS\,1257$-$326 from 2001 and early 2002. From
continued ATCA monitoring of the source we now have a number of
additional measurements of characteristic timescale, $t_{\rm scint}$.
For the present paper, we have reanalysed all data, taking $t_{\rm
scint}$ to be the half-width at $1/e$ of the intensity autocorrelation
functions. Note that this is slightly different from the half-width at
half-maximum definition used in Paper I. Furthermore, we have
recomputed the estimates of error in $t_{\rm scint}$, using the
formulation given in Appendix A of \citet{ric2002} to compute
estimation errors in the autocorrelations. Note that this reanalysis
makes no significant difference to any of the results presented in
Paper I. The updated measurements of $t_{\rm scint}$ are used and
shown in the following analysis section.

\section{Analysis}
\label{sec-analysis}
The time delay results can be combined with the annual cycle observed
in the variability timescale of PKS\,1257$-$326 to determine both the
scintillation velocity and the dimensions of the scintillation pattern.

The scintillation pattern is assumed to be elliptical with minor axis
scale length $a_{\rm min} = s_0 /\sqrt{R}$ and major axis scale $a_{\rm maj} =
s_0 \sqrt{R}$, and
is oriented along the vector $\hat{\bf S}=(\cos \beta, \sin \beta)$.
The scintillation velocity is written in the form ${\bf v}(T)=v_{\rm
ISS} - v_\oplus(T) \equiv(v_\alpha (T),v_\delta (T))$, and varies on
an annual cycle as a function of time, $T$, due to Earth's orbital
motion \citep{mj2002}.  Here we define $v_\oplus(T)$ as the
velocity of the Earth with respect to the solar system barycenter. 
In this notation, the time delay
expected between two telescopes displaced by a distance ${\bf r} =
(r_\alpha, r_\delta)$ is \citep{ck78},
\begin{eqnarray}
\Delta t = \frac{{\bf r} \cdot {\bf v} + (R^2 -1)({\bf r} \times
\hat{\bf S})({\bf v} \times \hat{\bf S} ) }{v^2 + (R^2-1)({\bf v}
\times \hat{\bf S})^2}, \label{TimeDelay}
\end{eqnarray}
where the dependence of ${\bf v}$ on $T$ is suppressed.  The
scintillation timescale is
\begin{eqnarray}
t_{\rm scint} = \frac{\sqrt{R} s_0}{\sqrt{v^2 + (R^2 -1) ({\bf v}
\times \hat{\bf S})^2}}. \label{TScint}
\end{eqnarray}
In practice neither the variability timescale nor the time delay
experiment datasets suffice to uniquely determine all five
scintillation parameters, $v_{{\rm ISS},\alpha}, v_{{\rm ISS},\delta},
\beta, R$ and $s_0$, at any one observing frequency, so we fit to both
the time delay and the annual cycle data simultaneously.  We also fit
both frequencies simultaneously, which imposes the additional
constraint that the scintillation velocity is identical at the two
frequencies; the anisotropic ratios of the scintillation patterns and
their associated position angles at the two frequencies are still
assumed independent.

Our fit minimises the function
\begin{eqnarray}
\chi^2 = \chi_{\rm \Delta t}^2 + \chi_{\rm t_{\rm scint}}^2,
\end{eqnarray}
where $\chi_{t_{\rm scint}}^2$ compares our $N_{t_{\rm scint}}=51$
annual cycle measurements, $m_{t_{\rm scint}}(T,\lambda)$, to a model,
$M_{t_{\rm scint}}(T,\lambda)$, based on equation (\ref{TScint})
\begin{eqnarray}
\chi_{t_{\rm scint}}^2 = \frac{1}{N_{t_{\rm scint}}} \sum_{\lambda}
\sum_{\rm T}  \frac{\left[ m_{t_{\rm scint}} (T,\lambda)-M_{t_{\rm
scint}} (T,\lambda)\right]^2}{ \sigma_{t_{\rm scint}}^2(T,\lambda) }.
\end{eqnarray}
The value of $\chi_{\Delta t}^2$ compares our $N_{\Delta t}=3$ time
delay measurements, $m_{\Delta t} (T,\lambda)$, to a model, $M_{\Delta
t} (T,\lambda)$, based on equation (\ref{TimeDelay}),
\begin{eqnarray}
\chi_{\Delta t}^2 = \frac{1}{N_{\Delta t}} \sum_{\lambda} \sum_{\rm T}
\frac{\left[ m_{\Delta t} (T,\lambda)-M_{\Delta t}
(T,\lambda)\right]^2}{ \sigma_{\Delta t}^2(T,\lambda) },
\end{eqnarray}
We have fitted only to the mean time delay from each observation, and
have not attempted to model the change in time delay due to the change
in telescope baseline during the course of the experiment.  As
discussed above in \S\ref{sec-results} we do not detect any
significant change in delay over the course of each observation.

The two contributions to $\chi^2$, $\chi_{t_{\rm scint}}^2$ and
$\chi_{\Delta t}^2$ are assumed independent, and are weighted so that
the relatively few time delay experiments are given equal importance
in the fit relative to the annual cycle.

In principle, this fitting procedure allows one to uniquely derive the
properties of the scintillation pattern and the peculiar velocity of
the scattering medium.  However, the system becomes degenerate as the
anisotropic ratio becomes large.  In the limit $R \rightarrow \infty$,
the time delay approaches $\Delta t = ({\bf r} \times \hat{\bf S}) /
({\bf v} \times \hat{\bf S})$ and the scintillation timescale
approaches $s_0 / \sqrt{R} ({\bf v} \times \hat{\bf S})$.  In this
limit one can solve for ${\bf v} \times \hat{\bf S}$ and $s_0 /
\sqrt{R}$, but not uniquely for $\beta$, and thus ${\bf v}_{\rm ISS}$.
Since the terms associated with anisotropy scale $\propto {\cal
O}(R^2)$ with respect to the isotropic terms in equations
(\ref{TimeDelay}) and (\ref{TScint}), the uniqueness of the solution
depends heavily on the anisotropy intrinsic to the scintillation
pattern. For $R \gtrsim 5$, there are degenerate solutions for the
scintillation velocity, although the scintillation length scale along
the minor axis can still be uniquely determined. To decouple the
errors we have fit for the major and minor axis scales $a_{\rm maj}$
and $a_{\rm min}$ rather than $s_0$ and $R$.  A selection of results
is shown in Table~\ref{tab-fit}.  In this table the position angle of
anisotropy is shown in degrees north through east to follow the usual
convention of radio image analysis, although the angle $\beta$ is
defined east through north. 

\begin{deluxetable}{crrrrrrrrr}
\tablecolumns{10}
\tabletypesize{\footnotesize}
\tablecaption{Scintillation parameters from fitting to the combined annual cycle and time delay measurements\label{tab-fit}}
\tablehead{
\colhead{} & \colhead{} & \colhead{} &
\multicolumn{3}{c}{---------~~4.9 GHz~~---------}
& \multicolumn{3}{c}{---------~~8.5 GHz~~---------} & \colhead{}\\
 \colhead{} & \colhead{$v_{{\rm ISS},\alpha}$} & \colhead{$v_{{\rm
  ISS},\delta}$} & \colhead{$a_{\rm min}$} & \colhead{$a_{\rm maj}$} &
  \colhead{$90-\beta$} & \colhead{$a_{\rm min}$} & \colhead{$a_{\rm
  maj}$} & \colhead{$90-\beta$} & \colhead{$\chi^2$} \\
 \colhead{} & \colhead{(km s$^{-1}$)} & \colhead{(km s$^{-1}$)} &
 \colhead{($10^4$ km)} & \colhead{($10^4$ km)} &
 \colhead{$^\circ($N$\rightarrow$E)} & \colhead{($10^4$ km)}
 & \colhead{($10^4$ km)} & \colhead{$^\circ($N$\rightarrow$E)} &
 \colhead{($N_{\rm DOF}$)}}
\startdata
(a) & 2.59 & 18.30 & 6.68 & 6.68 & 0     & 4.23 & 4.23 & 0     & 106 \\
$\sigma$\tablenotemark{a} 
   & 0.14 & 0.18  & 0.07 & 0.07 & fixed & 0.08 & 0.08 & fixed & (53) \\
\hline
(b) & $-256$ & 219 & 4.24 & 487 & $-52.8$ & 3.42 & 190 & $-52.6$ & 1.75 \\
$\sigma$\tablenotemark{b} 
    & 47     & 36  & 0.08 & 41  & 0.3     & 0.06 & 36  & 0.3     & (46) \\
\hline
(c) & 49.2 & $-11.5$ & 4.23 & 50.8 & $-54.8$ & 3.43 & 41.2 & $-54.6$ &
1.97 \\
$\sigma$\tablenotemark{c}
    & 7.1  & 5.1     & 0.08 & 0.1  & 0.4     & 0.06 & 0.1  & 0.5     &
(49) \\
\hline
(d) & 16.5  & 10.7  & 4.08 & 20.5 & $-54.1$ & 3.19 & 500 & $-50.0$ &
3.6 \\
$\sigma$\tablenotemark{d}
    & fixed & fixed & 0.07 & 1.0  & 0.4     & 0.05 & 390 & 0.3     &
(48) \\
\enddata
\tablenotetext{a}{Solution (a) is for an isotropic scintillation
  pattern, which is not consistent with the data.} 
\tablenotetext{b}{Solution (b) is for an anisotropic scintillation
  pattern. This solution produces a good fit to both the annual cycle
  and time delay data, however the scintillation velocity and the
  major axis scale of the scintillation pattern are poorly
  constrained.} 
\tablenotetext{c}{Solution (c) is found when the axial ratio of the
  pattern is constrained to be $R \leq 12$.} 
\tablenotetext{d}{Solution (d) is found when the scattering screen is fixed
  to the LSR velocity.} 
\tablecomments{For each solution, error estimates for all parameters
  are shown beneath the best fit values. The values for the global
  reduced $\chi^2$ and degrees of freedom ($N_{\rm DOF}$) for each fit
  are shown in the last column. See \S\ref{sec-analysis} for discussion.}
\end{deluxetable}

\begin{figure}
\epsscale{1} \plotone{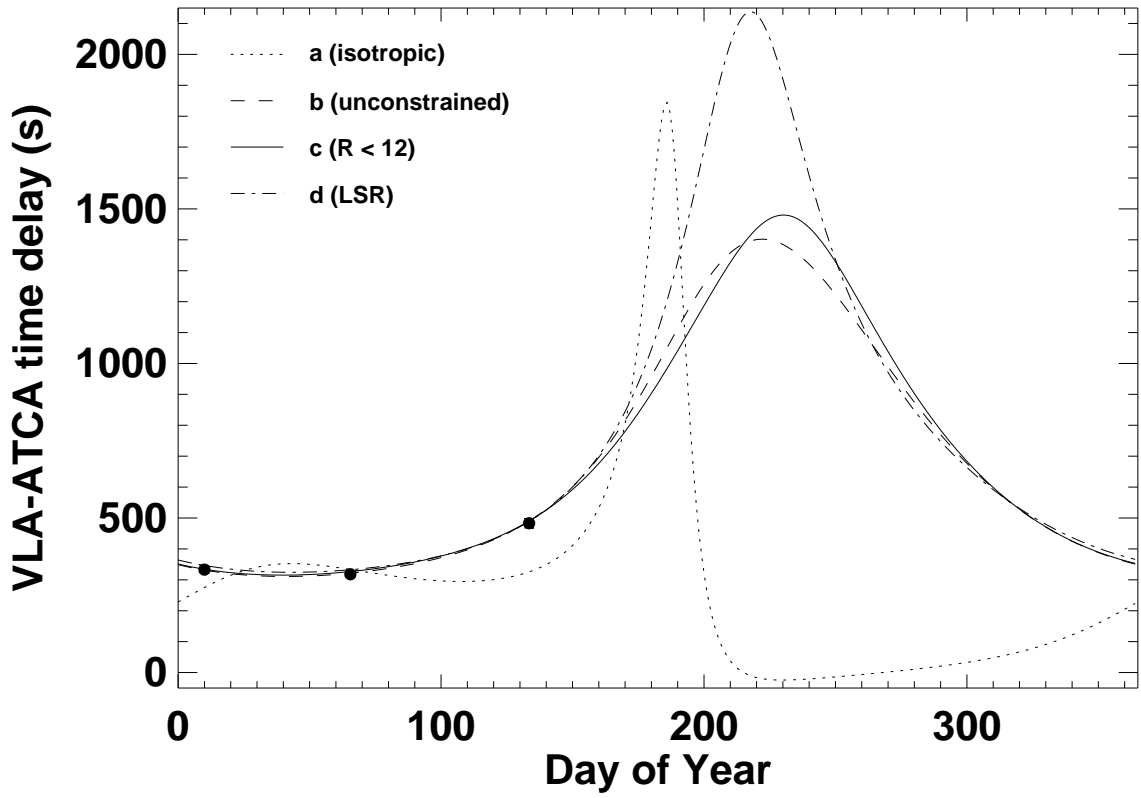}
\caption{Variation of time delay as a function of day of year for the
  solutions shown in Table~\ref{tab-fit}. Points near DOY 10, 67 and 134
  show the measured time delays.}
\label{fig-tdmodel}
\end{figure}

\begin{figure}
\epsscale{1} \plotone{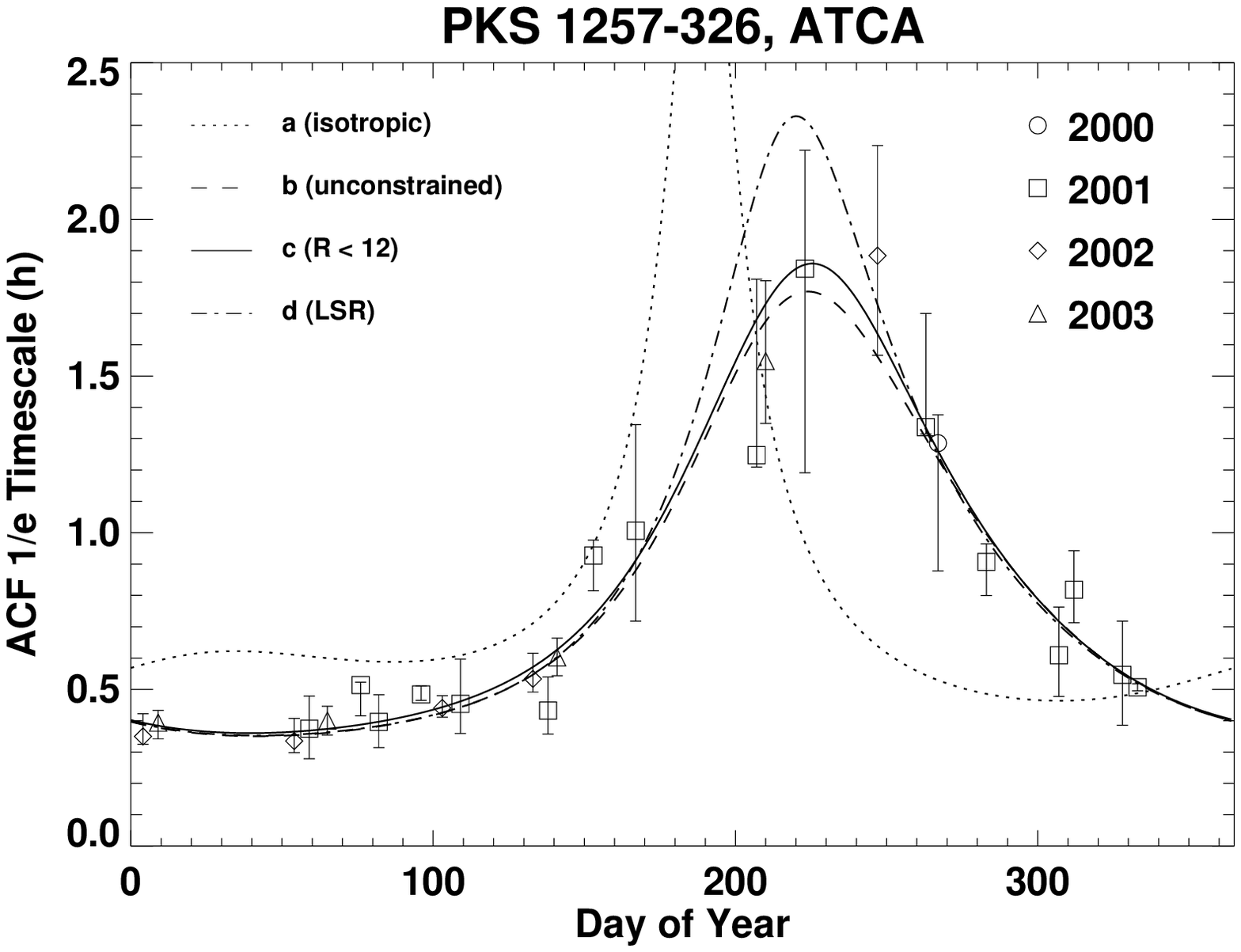}
\caption{Annual cycle in characteristic timescale, $t_{\rm scint}$,
  observed in ATCA data at 4.8\,GHz (see \S\ref{sec-ac}). The plotted
  lines show the model annual cycles for solutions in
  Table~\ref{tab-fit}.}
\label{fig-acfit}
\end{figure}

The time delays and characteristic timescale at 4.9\,GHz are plotted
as a function of day of year in Figure~\ref{fig-tdmodel} and
Figure~\ref{fig-acfit}, respectively.  Model predictions of each of
the solutions presented in Table~\ref{tab-fit} are shown along with
the observational data points.

The annual cycle and time delay measurements were found to be
inconsistent with a solution assuming an isotropic scintillation
pattern.  In particular, there are {\em no} solutions for the
isotropic case that fit time delays anywhere near as large as those
observed in 2002 May and also match the observed ``slow-down'' in
characteristic timescale which peaks around August. With anisotropy
allowed, solutions can be obtained which fit the observations well.
There are degenerate solutions for the scintillation velocity,
however, as $v_{{\rm ISS},\alpha}$ and $v_{{\rm ISS},\delta}$ are
strongly correlated. The best fit velocities tend to lie along a
straight line in the $(v_{{\rm ISS},\alpha},v_{{\rm ISS},\delta})$
plane.  The major axis scale is also poorly constrained by the annual
cycle and time delay data alone.

Another observational constraint on $a_{\rm maj}$ comes from the
spatial decorrelation of the pattern between the two telescopes.  To
model the expected decorrelation we assume that the intensity
autocorrelation decreases quadratically from unity near the origin,
which is a good approximation if the source has a gaussian brightness
distribution and source structure determines the size of the
scintillation pattern.  We find that when the pattern is highly
anisotropic and large compared to the baseline $|{\bf r}|$, the degree
of cross-correlation of the two observed patterns is approximately
\begin{eqnarray}
\rho({\bf r},\Delta t) = 1 - \frac{({\bf r} \times {\bf
  v})^2}{a_{\rm maj}^2({\bf v} \times \hat{\bf S})^2}.
\label{eq-decor}
\end{eqnarray}

By far the longest time delays were observed in the May epoch, as in
May the scintillation velocity has a large component perpendicular to
the baseline ${\bf r}$ as well as a large component parallel to the
major axis of the scintillation pattern, along $\hat{\bf S}$.  This
geometry also implies that we would expect to observe the largest
decorrelation of the scintillation pattern in May (see
Figure~\ref{fig-geom} and equation~\ref{eq-decor}).  Indeed $\rho({\bf
r},\Delta t)$ was observed to be somewhat smaller in May than in the
other two epochs. Although $\rho$ is dependent on the velocity, the
measured values of $\rho$ in any case provide a lower limit on $a_{\rm
maj}$. The lower limit is derived from the model which has the
smallest component of screen velocity parallel to $a_{\rm maj}$.  For
a scattering screen moving with the local standard of rest (solution
(d) in Table~\ref{tab-fit}), $a_{\rm maj}$ is found to be $\sim
5\times 10^5$\,km at 4.9~GHz (a factor of 2 smaller at 8.5\,GHz, but
the measurement errors are smaller at 4.9\,GHz).  Better fits to the
annual cycle data are found for models with somewhat larger velocity
components parallel to $a_{\rm maj}$.  These larger velocities require
an even larger $a_{\rm maj}$, i.e. even more extreme anisotropy in the
scintillation pattern. For example, solution (b) in
Table~\ref{tab-fit} implies a characteristic major axis scale of more
than $4\times 10^6$~km, and solution (c) implies $a_{\rm maj} \approx
1.2\times 10^6$~km at 4.9\,GHz.  If there is a systematic bias in the
determination of $\rho$, this should cancel out by taking the ratio of
$\rho$ for different epochs in equation~\ref{eq-decor} above.
Determining $a_{\rm maj}$ in this way was found to give essentially
the same results as using $\rho$ for the May epoch alone in 
equation~\ref{eq-decor}. Combining the results with the constraint
that $a_{\rm min} = 4.2 \times 10^4$~km at 4.9\,GHz implies a lower
limit for the pattern axial ratio $R \gtrsim 12$.

As discussed in Paper I, scintillations observed in PKS~1257$-$326 at
frequencies of 4.8\,GHz and higher are evidently in the weak
scattering regime, where the scintillation length scale is set by
whichever is the larger of the angular Fresnel scale, $\theta_{\rm F}
= \sqrt{c/(2\pi\nu L)}$, and the source angular size.  We can relate
the Fresnel radius, $r_{\rm F}= L\theta_{\rm F}$, to $a_{\rm min}$,
the $1/e$ value of the intensity autocorrelation function along the
minor axis, to find upper limits for the screen distance $L$.  For a
point source and an isotropic scattering screen, $a_{\rm min}$
corresponds to $1.2 r_{\rm F}$, yielding a direct measure of $L$. If
source structure influences the scintillation pattern, then
$a_{\rm min} > 1.2 r_{\rm F}$, giving an upper limit for $L$. For an
anisotropic scattering medium, the $1/e$ decorrelation point
asymptotes to $0.78 r_{\rm F}$ along the short axis, and the
corresponding upper limit for $L$ is given by $a_{\rm min} \geq 0.78
r_{\rm F}$. Uncertainties in $L$ derived this way scale as the square
of uncertainties in $a_{\rm min}$. From the best fits to the
data, $a_{\rm min}$ is well-determined to within $\sim 2$\% despite
the degeneracies in velocity and pattern axial ratio. At 4.9\,GHz,
$a_{\rm min} = 4.2\pm 0.1\times 10^4$~km gives $L \leq 4$\,pc if the
scattering medium is isotropic. For an anisotropic scattering
medium, incorporating the uncertainty in $a_{\rm min}$, this limit
becomes $L \leq 10$\,pc.

\section{Discussion and conclusions}

\subsection{Anisotropy in the scintillation pattern}

\begin{figure}
\epsscale{1}\plotone{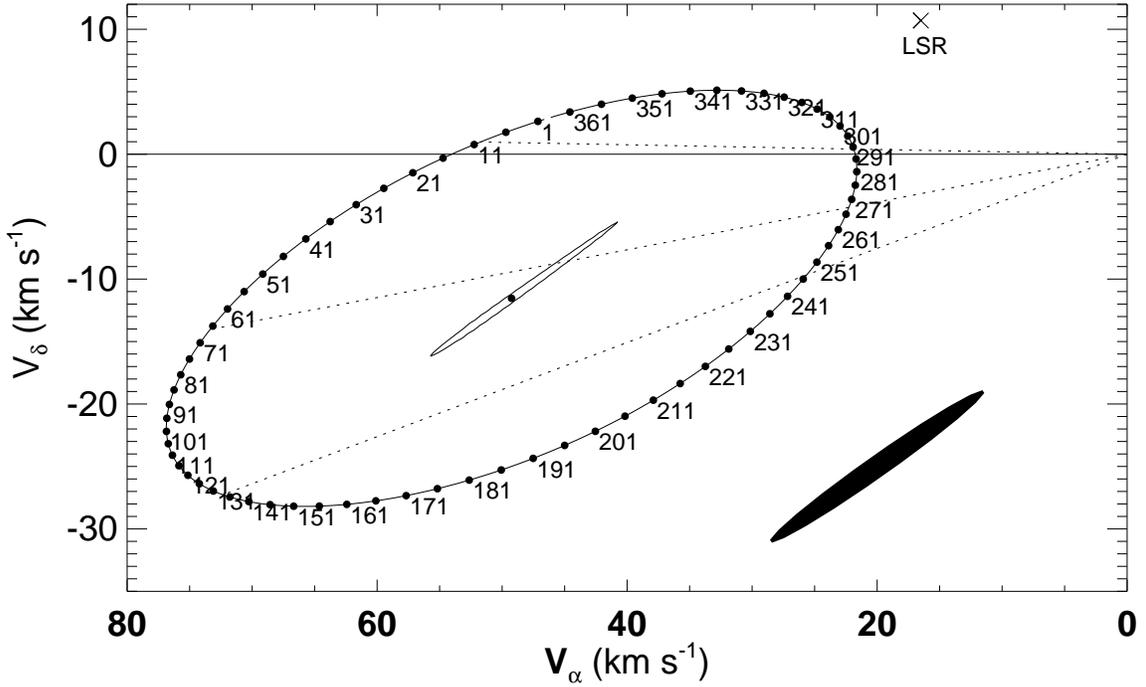}
\caption{Scintillation velocity projected onto the plane of the sky,
 showing the annual variation of the Earth's velocity with respect to
 the scattering screen for anisotropic solution (c) in
 Table~\ref{tab-fit}. 
 Dotted lines show the velocity on the days of
 the time delay observations. The filled ellipse represents a
 contour of the best-fit scintillation pattern at 4.9\,GHz. The
 1-$\sigma$ contour of the best fit solution for $v_{{\rm
 ISS},\alpha}$ versus $v_{{\rm ISS},\delta}$ is plotted around the
 best fit velocity. The component of velocity parallel to the long
 axis of the scintillation pattern is poorly constrained.
 If the scattering screen were moving with
 the local standard of rest, the center of the ellipse would be at
 position marked in the top right-hand corner.}
\label{fig-ell}
\end{figure}

Figure~\ref{fig-ell} shows the scintillation velocity over the course
of the year for PKS~1257$-$326 along with a contour of the fitted
scintillation pattern projected on the plane of the sky for
anisotropic solution (c) in Table~\ref{tab-fit}. Note that in Paper I
the geometry (Figure 4 in Paper I) was displayed as if looking down
on the Earth whereas in the present paper we show the view looking up
at the sky, corresponding to a reflection of the east-west axis. It is
interesting to note that the long axis of the scintillation pattern is
aligned in a north-west direction.  This is close to the direction of
the main extended arcsecond-scale jet in the source, and also with the
milliarcsecond-scale jet components observed in VLBI data (Bignall et
al., in preparation).  \citet{ric2002} argued that for the case of
PKS~0405$-$385 there is strong evidence that the scintillations result
from anisotropic scattering in a thin screen. Likewise for
PKS~1257$-$326 we observe the same large negative ``overshoot'' in the
light-curve autocorrelations as investigated by \citet{ric2002} and
shown to be due to anisotropic scattering. 
However, given the coincidence of position angle, it
may be the case for PKS~1257$-$326 that the anisotropy in the
scintillation pattern is also partly related to elongation of 
the source itself. It is possible that if the source were elongated in
a different direction, then such large and rapid scintillation might
not be observed through the anisotropic scattering screen. Further
analysis is required before conclusions can be drawn as to whether the
anisotropy is entirely due to the scattering screen, source structure,
or a combination of both source and screen.

The origin of the anisotropy may be resolved through a detailed
analysis of the polarized flux density scintillations which is
ongoing. The ATCA observations provide the full set of Stokes
parameters, $I$, $Q$, $U$ and $V$. While no significant circular
polarization has been detected for PKS 1257$-$326 ($3\sigma$ upper
limit $V = 0.2$\,mJy, or 0.1\% of $I$), we detect significant flux
density in the linear polarization parameters $Q$ and $U$.  If
anisotropy is a property of the scattering medium, then the same
degree of anisotropy should be present in all Stokes parameters, while
if a property of the source, then the linear polarization is not
likely to have the same anisotropy as Stokes $I$.  Cross-correlations
between the  various Stokes parameter pairs can also be used to reveal
the microarcsecond-scale polarized structure of the source
\citep{mj2002,ric2002}.  More information on the properties of both
source and scatterer can also be obtained through a detailed analysis
of light curve structure functions and power spectra -- here we have
used only the $1/e$ decorrelation timescale to characterise the
scintillations.

\subsection{Implications for the source}

For a scattering screen at $L=10$\,pc, the angular scale of the
scintillation pattern implies that the angular size of the
scintillating component of the source cannot be much larger than 
$\sim 300$ by 30 $\mu$as at 4.9\,GHz (a factor $\sim 1.2$ smaller at
8.5\,GHz).  Based on the flux density in the unresolved VLBI core, the
observed modulation index and also the asymmetry coefficient analysis
of \citet{sst2005}, we estimate that the flux density of the scintillating
component is close to 100\,mJy at 4.9\,GHz. 
Then the lower limit for the implied brightness temperature
of this component is $T_b \approx 1.4\times 10^{12}$~K. If the
anisotropy is caused purely by the scattering medium and the source is
intrinsically circularly symmetric with angular diameter 30\,$\mu$as,
then the implied brightness temperature becomes a factor of 10
higher. On the other hand, if the scattering screen is in fact closer
than 10\,pc, this implies a larger angular size and hence lower $T_b$
for the source.  $T_b \sim 10^{13}$~K is still within the range
inferred from VLBI observations of other flat-spectrum quasars
\citep[e.g.][]{hor2004}.  A Doppler factor of $\sim 10$--20 is implied
to reduce the intrinsic brightness temperature below the inverse
Compton limit, however the relativistic beaming evidently does not
correspond to an expansion of the emitting region. While there has
been some slow evolution of the source flux density over three years
of monitoring, we observed a repeated annual cycle, 
and rapid scintillation has been seen every time
PKS~1257$-$326 has been observed since 1995, indicating that the
compact component is long-lived and relatively stable over at least a
decade.

\subsection{Implications for the scattering plasma}

The MASIV Survey showed that such extreme intra-hour variability (IHV)
as observed in PKS~1257$-$326, PKS~0405$-$385 and J1819+3845 is
exhibited by $\ll 1$\% of compact flat-spectrum radio sources.
Combined with the interpretation of the rapid variability as due to
scattering in the very local ISM, this suggests that the covering
fraction of such nearby scattering plasma is very
small. Alternatively, it is possible that a large fraction of sources
may be sufficiently scatter-broadened by the more distant ISM to
quench the weak scintillations in more nearby scattering plasma.
Further investigation of these possibilities and of the local ISM
structure inferred from other observations is needed to fully
understand the IHV phenomenon. For a transition frequency between
strong and weak scattering of $\sim 5$\,GHz, as indicated by the
observed peak of modulation close to this frequency, and a screen
distance of 10\,pc, the scattering measure ($SM$) in conventional
units of kpc m$^{-20/3}$ is found to be $\log SM \sim -2.8$ using
equation 17 from \citet{cl2002}.

All three fast scintillators show evidence of highly anisotropic
scintillation patterns \citep{ric2002,dtdb2003}. Interestingly, highly
anisotropic scattering has also been inferred in the interpretation of
parabolic arcs observed in the secondary spectra of pulsars
\citep[e.g.][]{wal2004}. \citet{rc2004} discuss the evidence for
highly anisotropic plasma turbulence concentrated in relatively thin
layers, suggesting filamentary density structures aligned by the local
magnetic field. If the scattering occurred in a very extended medium,
the magnetic field, and hence anisotropy in the electron density
fluctuations, would tend to have different orientations through the
scattering medium and thus the scintillation pattern would be expected
to be more isotropic.

\subsection{Conclusions}

We have presented a combined analysis of two-station pattern arrival
time delays and annual cycles in the scintillation timescale for the
quasar PKS~1257$-$326. This is the first time such a combined fitting
method has been applied to data on a scintillating quasar. We conclude
that: 
\begin{itemize}

\item the large, rapid flux density variations observed in PKS~1257$-$326
are entirely due to interstellar scintillation.

\item the scattering occurs in a confined region within 10\,pc of the
  solar system.

\end{itemize}

The scintillation  velocity and geometry of the scintillation pattern
for PKS 1257$-$326 could not be uniquely determined using the annual
cycle data alone (Paper I), although none of the conclusions presented
in Paper I are  significantly altered by the new results.
Analysis of the combined annual cycle and time delay data shows that
there is a high degree of anisotropy in the scintillation pattern. The
characteristic linear scale of the pattern along its short axis is
well determined, as is the position angle along which the
scintillation pattern is elongated.  The large anisotropy results in a
degeneracy in fitting for the scattering screen velocity, with the
best fit velocities  lying along a line in the
$(v_{\alpha},v_{\delta})$ plane. The scintillation scale on the long
axis is constrained using measurements of the spatial decorrelation of
the pattern between the VLA and ATCA.  An axial ratio of at least 12:1
is inferred. 

The implied source brightness temperature is high, up to $10^{13}$\,K,
but not incompatible with brightness temperatures derived from
VLBI observations of other compact quasars. Although a compact source
is a requirement for ISS to be observed, it seems likely that it is
the unusual scattering properties of the local ISM on the
line-of-sight to PKS~1257$-$326, rather than an intrinsic property of
the source itself, which give rise to the unusually rapid
scintillation observed in this case.

Pattern arrival time delays between widely separated telescopes,
combined with measurements of the scintillation timescale, provide a
powerful method of determining scintillation parameters for
quasars which scintillate rapidly. In order to measure such time
delays however, a source must vary by a detectable amount (typically
of order 1 mJy or more) over a period of order 1 minute.  Such
variability is observed in very few sources. Another problem is that
the scintillation behaviour for some sources is episodic, making it
more difficult to coordinate observations involving multiple
telescopes.  With more sensitive telescopes coming on line in the
future, one would expect to find more low flux density rapid
scintillators, thus probing scattering on many more lines of sight
through the local ISM.

\begin{acknowledgements}
We thank Barney Rickett for valuable discussions, and an anonymous
referee for their helpful suggestions to improve the paper. 
The ATCA is part of the Australia Telescope, which is funded by the
Commonwealth of Australia for operation as a National Facility managed
by CSIRO. The VLA is operated by the National Radio Astronomy
Observatory, a facility of the National Science Foundation operated
under cooperative agreement by Associated Universities, Inc. 
\end{acknowledgements}

\bibliographystyle{apj} 
\bibliography{all}

\end{document}